\documentclass[twocolumn]{author}
\usepackage{graphicx, color, soul}
\usepackage{caption, multirow}
\usepackage[colorlinks, linkcolor=black, citecolor=cyan]{hyperref} 

\hfuzz=\maxdimen

\newcommand{\Q}[1]{`#1'}
\newcommand{\fig}[1]{\mbox{Figure \ref{#1}}}
\newcommand{\tab}[1]{\mbox{Table \ref{#1}}}

\newcommand{\red}{\textcolor{black}}

\usepackage[acronym]{glossaries}
\setacronymstyle{long-short}
\newacronym{AUC}{AUC}{Area Under Curve}
\newacronym{ANOVA}{ANOVA}{Analysis of Variance}
\newacronym{CI}{CI}{Confidence Interval}
\newacronym{CNN}{CNN}{Convolution Neural Network}
\newacronym{eTSA}{eTSA}{endoscopic TransSphenoidal Approach}
\newacronym{ECC}{ECC}{Enhanced Correlation Coefficient maximisation}
\newacronym{EMA}{EMA}{Exponential Moving Average}
\newacronym{EPS}{EPS}{Endoscopic Pituitary Surgery}
\newacronym{FP}{FP}{Floating Point}
\newacronym{FPS}{FPS}{frames per second}
\newacronym{IoU}{IoU}{Intersection over Union}
\newacronym{IQR}{IQR}{Inter-Quartile Range}
\newacronym{LSTM}{LSTM}{Long Short-Term Memory}
\newacronym{mAP}{mAP}{mean average precision}
\newacronym{mIoU}{mIoU}{mean IoU}
\newacronym{MLP}{MLP}{MultiLayer Perceptron}
\newacronym{mOSATS}{mOSATS}{modified OSATS}
\newacronym{MOTA}{MOTA}{Multiple Object Tracking Accuracy}
\newacronym{MOTP}{MOTP}{Multiple Object Tracking Precision}
\newacronym{NSAK}{NSAK}{Noise Scale Adaptive Kalman}
\newacronym{OSATS}{OSATS}{Objective Structured Assessment of Technical Skills}
\newacronym{PRINTNet}{PRINTNet}{Pituitary Real-time INstrument Tracking Network}
\newacronym{PCC}{PCC}{Pearson Correlation Coefficient}
\newacronym{RF}{RF}{Random Forest}
\newacronym{SORT}{SORT}{Simple Online and Realtime Tracking}
\newacronym{SVM}{SVM}{Support Vector Machine}

\newacronym{CRUK}{CRUK}{Cancer Research UK}
\newacronym{DSIT}{DSIT}{Department of Science, Innovation and Technology}
\newacronym{EPSRC}{EPSRC}{Engineering and Physical Sciences Research Council}
\newacronym{IRB}{IRB}{Institutional Review Board}
\newacronym{NHNN}{NHNN}{National Hospital for Neurology and Neurosurgery}
\newacronym{NIHR}{NIHR}{National Institute for Health and Care Research}
\newacronym{UCL}{UCL}{University College London}
\newacronym{WEISS}{WEISS}{Wellcome/EPSRC Centre for Interventional and Surgical Sciences}

\begin{document}

\title{Automated Surgical Skill Assessment in Endoscopic Pituitary Surgery using Real-time Instrument Tracking on a High-fidelity Bench-top Phantom}

\author{
\mbox{Adrito Das$^{1}$},
\mbox{Bilal Sidiqi$^{1}$},
\mbox{Laurent Mennillo$^{1}$},
\mbox{Zhehua Mao$^{1}$},
\mbox{Mikael Brudfors}$^{2}$,
\mbox{Miguel Xochicale$^{1,3}$},
\mbox{Danyal Z. Khan$^{1,4}$},
\mbox{Nicola Newall$^{1,4}$},
\mbox{John G. Hanrahan$^{1,4}$},
\mbox{Matthew J. Clarkson$^{1,5}$},
\mbox{Danail Stoyanov$^{1}$},
\mbox{Hani J. Marcus$^{1,4}$},
and 
\mbox{Sophia Bano$^{1}$}
}
\address{
$^{1}$ Wellcome / EPSRC Centre for Interventional and Surgical Sciences, University College London, London, UK\\
$^{2}$ NVIDIA\\
$^{3}$ School of Biomedical Engineering and Imaging Sciences, King’s College London, London, UK\\
$^{4}$ Department of Neurosurgery, National Hospital for Neurology and Neurosurgery, London, UK \\
$^{5}$ Department of Medical Physics \& Biomedical Engineering, University College London, London, UK
}

\historydate{Pre-print: 25 September 2024}

\abstract
{Improved surgical skill is generally associated with improved patient outcomes, although assessment is subjective; labour-intensive; and requires domain specific expertise. Automated data driven metrics can alleviate these difficulties, as demonstrated by existing machine learning instrument tracking models in minimally invasive surgery. However, these models have been tested on limited datasets of laparoscopic surgery, with a focus on isolated tasks and robotic surgery. In this paper, a new public dataset is introduced, focusing on simulated surgery, using the nasal phase of endoscopic pituitary surgery as an exemplar. Simulated surgery allows for a realistic yet repeatable environment, meaning the insights gained from automated assessment can be used by novice surgeons to hone their skills on the simulator before moving to real surgery. PRINTNet (Pituitary Real-time INstrument Tracking Network) has been created as a baseline model for this automated assessment. Consisting of DeepLabV3 for classification and segmentation; StrongSORT for tracking; and the NVIDIA Holoscan SDK for real-time performance, PRINTNet achieved 71.9\% Multiple Object Tracking Precision running at 22 Frames Per Second. Using this tracking output, a Multilayer Perceptron achieved 87\% accuracy in predicting surgical skill level (novice or expert), with the \Q{ratio of total procedure time to instrument visible time} correlated with higher surgical skill. This therefore demonstrates the feasibility of automated surgical skill assessment in simulated endoscopic pituitary surgery. The new publicly available dataset can be found here: \textcolor{blue}{https://doi.org/10.5522/04/26511049}.}

\maketitle

\section{Introduction}

\begin{figure*}[t]
\centering
\includegraphics[width=0.99\textwidth]{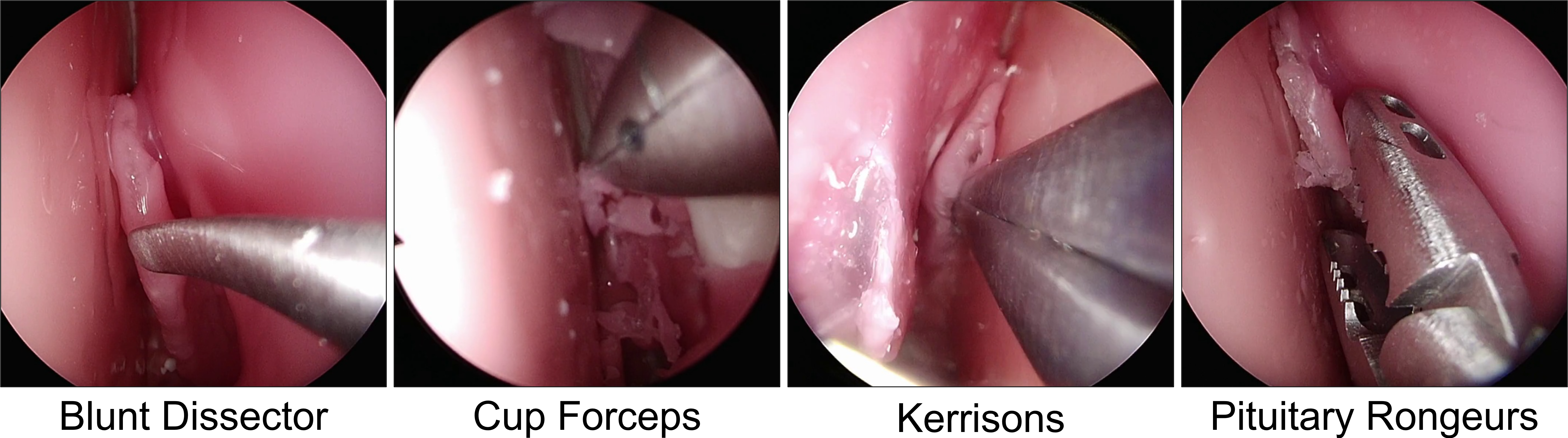}
\caption{Representative images of the 4-instrument-classes used in the nasal phase of endoscopic pituitary surgery.}
\label{fig1}
\end{figure*}

\begin{figure*}[t]
\centering
\includegraphics[width=0.99\textwidth]{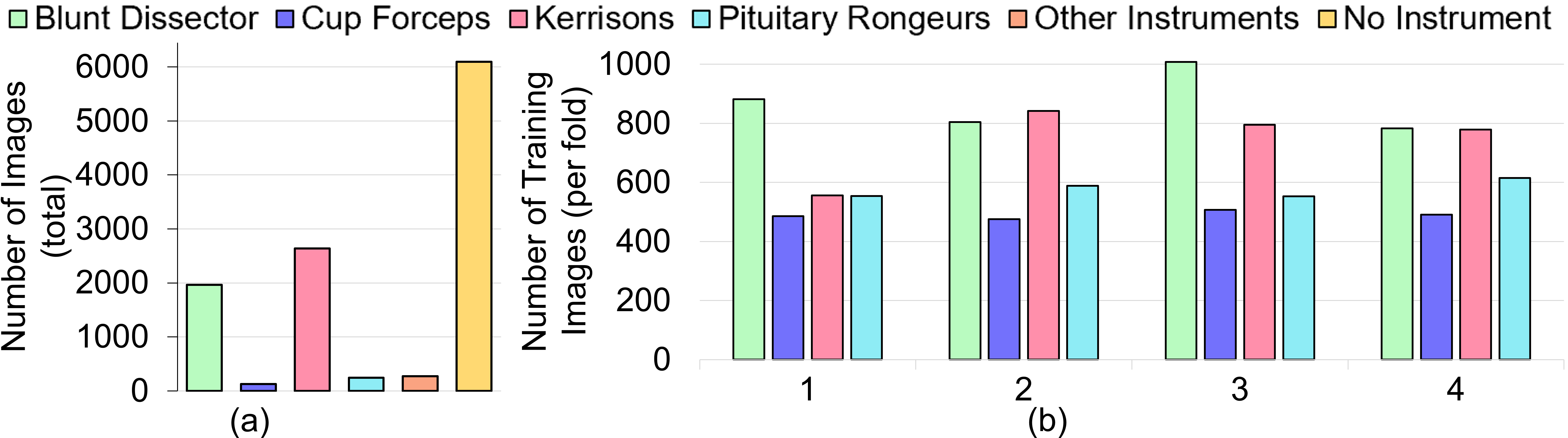}
\caption{Distribution of instruments: (a) Total number of images before data balancing; (b) Number of images per fold after data balancing.}
\label{fig2}
\end{figure*}

Benign tumours of the pituitary gland, pituitary adenomas, are common, associated with systemic morbidity and mortality, and the majority are curable with surgery \cite{Asa2008}\cite{Ezzat2004}\cite{Khan2023}. The \gls{eTSA}, is a minimally invasive surgery where these tumours are removed by entering through a nostril \cite{Marcus2021}. However, this surgery has a steep learning curve, with superior surgical skill generally associated with superior patient outcomes \cite{Khan2023}\cite{Leach2010}\cite{McLaughlin2012}.

\gls{OSATS} measures surgical skill by assessing how well aspects of a surgical task are performed on a scale of 1-5 \cite{Martin1997}. For example, for the aspect of instrument handling, a value of 1 indicates \Q{Repeatedly makes tentative or awkward moves with instruments}, and a value of 5 indicates \Q{Fluid moves with instruments and no awkwardness} \cite{Newall2022}. However, it is not operation specific; liable to interpreter variability; and is a time-consuming manual process requiring surgical experts \cite{Niitsu2012, Khan2024A}. Data driven metrics may be more specific; objective; reproducible; and easier to automate. 

Neural networks can automatically and accurately determine surgical skill \cite{Pedrett2023, Khan2024B}. More specifically, instrument tracking has been shown to be associated with \gls{OSATS} in minimally invasive surgeries \cite{Pedrett2023}. However, the models have been tested on limited datasets with a focus on laparoscopic surgeries \cite{Pedrett2023}. Pedrett et al. \cite{Pedrett2023} provides a comprehensive list of these datasets, which are videos of: isolated tasks (e.g. peg transfers in JIGSAWS \cite{JIGSAWS}); real surgery (e.g. Cholec-80 with no publicly available surgical skill assessment \cite{Lavanchy2021}); on robotic surgery (e.g. ROSMA \cite{ROSMA}); or include instrument tracking data from built in methods (e.g. \cite{PrezEscamirosa2019}) or wearable sensors (e.g. \cite{Soangra2022}). 

In this paper, this previous work is extended to be tested on videos of a high-fidelity bench-top phantom of the full nasal phase of \gls{eTSA}. These videos are therefore of a non-laparoscopic; non-private; non-robotic; and non-task-isolated surgery with no tracking data. This phantom is commonly used in neurosurgical training to simulate real surgery, and so surgical skill is an important measure to track a novice surgeon’s progress until they are able to perform real surgery. Additionally, the insights gained from the automated assessment can be used to isolate specific areas of improvement for the novice surgeon. In real surgery, surgeons are already of sufficient skill, and surgical skill assessment has the alternative use of correlating certain practices with patient outcomes. 

Moreover, instrument tracking in \gls{eTSA} provides a unique computer vision challenge due to: (I) A non-fixed endoscope leading to large camera movements; (II) The frequent withdrawal of instruments leading to instruments having a range of sizes; (III) The use of niche instruments leading to heavy class imbalance; (IV) The smaller working space requiring the use of a wide lens, distorting images (see \fig{fig1}). To overcome these challenges, \gls{PRINTNet} has been created, and the output is used to demonstrate correlations between instrument tracking and surgical skill. Therefore, this paper's contribution are:

\begin{enumerate}
    \item The first public dataset containing both instrument and surgical skill assessment annotations in a high-fidelity bench-top phantom of \gls{eTSA}.
    \item A baseline network capable of automated classification; segmentation; and tracking of the instruments in the nasal phase of \gls{eTSA}, integrated on a NVIDIA Clara AGX for real-time assistance in surgical training sessions. 
    \item Statistical analysis between instrument tracking and surgical skill assessment in \gls{eTSA}.
\end{enumerate}

\section{Related work}\label{sec2}
Instrument classification in \gls{eTSA} has been attempted in the PitVis-EndoVis MICCAI-2023 sub-challenge \cite{Das2024}, where 25-videos and 8-videos of real \gls{eTSA} (complete videos) were used for training and testing respectively.

Instrument segmentation and tracking is yet to be explored for \gls{eTSA}, though it has been attempted in minimally invasive surgeries since 2016 \cite{Rueckert2024}\cite{Wang2022}. Modern models use encoder-decoder architectures, utilising U-Net \cite{Ronneberger2015} and its variants for segmentation \cite{Rueckert2024}, and early forms of SORT \cite{Bewley2016} for tracking \cite{Wang2022}. 

The most similar study to this paper linking instrument tracking to surgical skill assessment is one conducted on robotic thyroid surgery \cite{Lee2020}. 23-videos (simulation and real) were used for training the 4-instrument-class tracking model, and 40-simulation-videos were used for training the surgical assessment model, with 12-simulation-videos used for testing \cite{Lee2020}. Mask R-CNN and DeepSORT were used for segmentation and tracking respectively, achieving 70.0\% \gls{AUC} for tracking a tool tip within 1mm \cite{Lee2020}. A \gls{RF} model was shown to be the best predictor of surgical skill, achieving 83\% accuracy in distinguishing between novice; intermediate; and expert surgeons \cite{Lee2020}. It was found \Q{economy of motion} was the most important predictive factor in where camera motion is minimal \cite{Lee2020}.

Other studies that use tool tracking for surgical skill assessment include one on real non-robotic laparoscopic cholecystectomy \cite{Fathollahi2022}. Here, instruments in 80-videos (15-test) of the calot triangle dissection phase in were tracked \cite{Fathollahi2022}. The model consisted of YoloV5 for detection, followed by a Kalman filter and the Hungarian algorithm for tracking, achieving 83\% \gls{MOTA} and 83\% accuracy in binary skill assessment via \gls{RF} \cite{Fathollahi2022}. Alternative models, such as those utilising aggregation of local features, have also been used \cite{Li2022}. This model consisted of stacked \glspl{CNN} followed by bidirectional \glspl{LSTM} and temporal pooling \cite{Li2022}. On 24-videos (4-fold) of the calot triangle and gallbladder dissection phases of real non-robotic laparoscopic cholecystectomy the model achieved 46\% Spearman’s rank correlation on a 1-5 scale \cite{Li2022}. An identical model trained on 30-videos (4-fold) of the 3 isolated robotic tasks found in the JIGSAWS dataset \cite{JIGSAWS} achieved 83\% Spearman’s rank correlation on a 1-6 scale \cite{Li2022}. This paper extends these methods to a new and unique dataset, in order to test their capability.

\section{Dataset description}

\subsection{Videos}
During a surgical training course at \red{the National Hospital for Neurology and Neurosurgery, London, UK}, 15 simulated surgeries videos (11426-images) were recorded, one per participating surgeon, using a commercially available high-fidelity bench-top phantom of the nasal phase of \gls{eTSA}\footnote{www.store.upsurgeon.com/products/tnsbox/} \cite{Newall2022}. 
The participants were recruited from multiple neurosurgical centres within \red{the United Kingdom}, with self-reported skill levels (10-novice, 5-expert), receiving tutorials and teaching beforehand. A high-definition endoscope (\red{Olympus S200 visera elite endoscope}) was used to record the surgeries at 25-\gls{FPS} with $720\times1080$ resolution, and stored as .mp4 files in a \red{surgical video management and analytics platform (Medtronic, Touch Surgery\textsuperscript{TM} Ecosystem\footnote{https://www.touchsurgery.com/})}. Ethical approval was granted \red{by the \gls{IRB} at \gls{UCL} (17819/011)} with informed participation consent.
 
\subsection{Instrument annotations}
Each video was sampled at 1-\gls{FPS} with $720\times1080$ resolution, and stored as .png files. Third party annotators (\red{Anolytics\footnote{https://www.anolytics.ai/}}) manually annotated each image for instrument boundary and class, which was then verified by two neurosurgical trainees and one consultant neurosurgeon. No image contained multiple instruments, and only visible parts of the instrument were annotated if obscured. \fig{fig2}a displays the distribution of the instruments.

\subsection{Surgical skill assessments}
\begin{figure*}[t]
\centering
\includegraphics[width=0.99\textwidth]{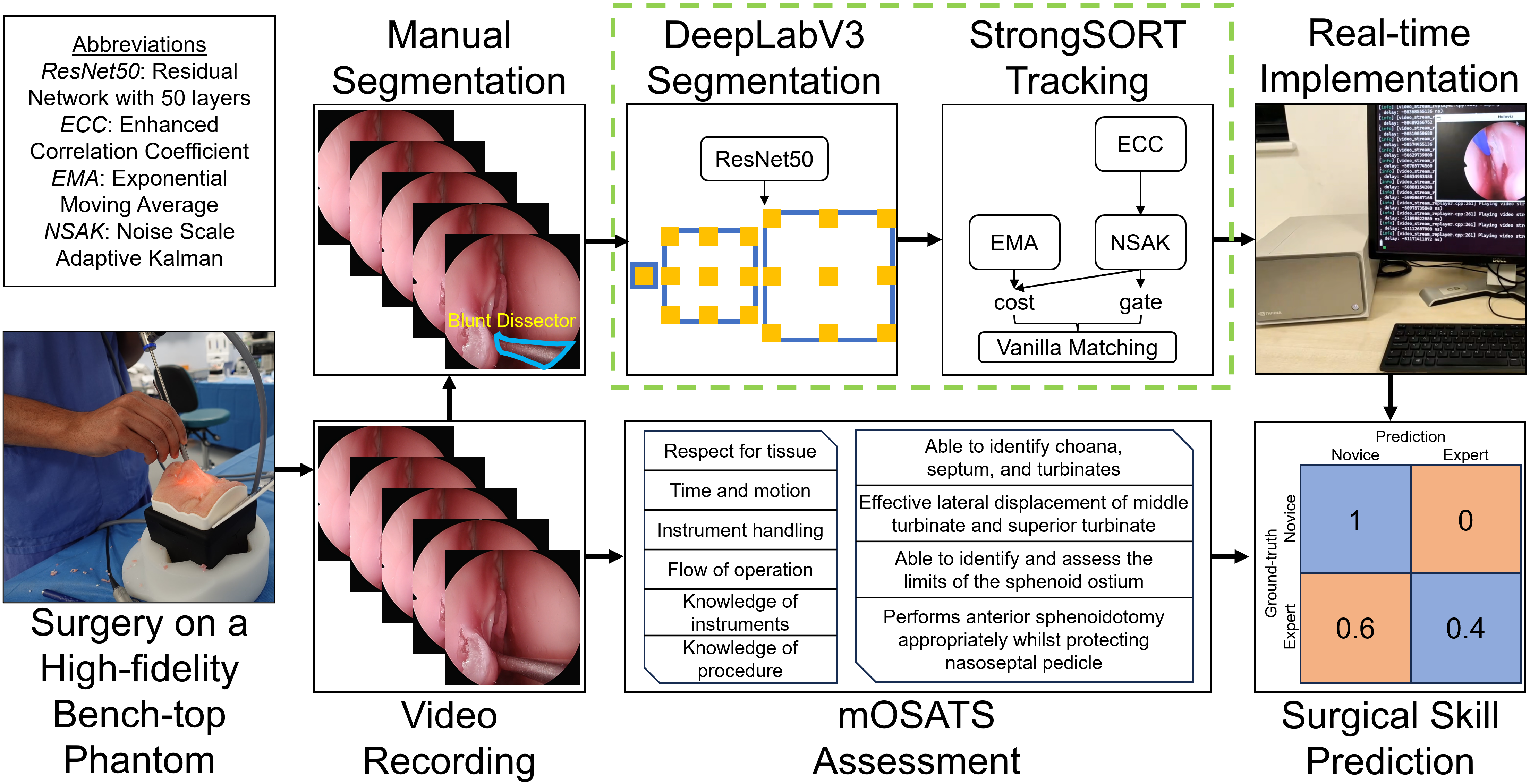}
\caption{Complete workflow diagram of this study.}
\label{fig4}
\end{figure*}

\begin{figure}[t]
\centering
\includegraphics[width=0.45\textwidth]{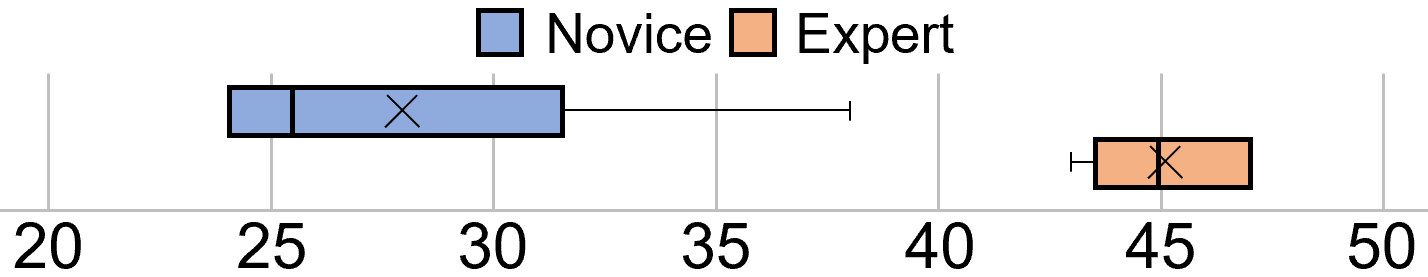}
\caption{Distribution of mOSATS (10-aspects, max 50) across the 15-videos.}
\label{fig3}
\end{figure}

\Gls{mOSATS}, \gls{OSATS} curated for pituitary videos, was created, leading to 10-aspects each measured between 1-5 \cite{Newall2022}. Each video was assessed by two neurosurgical trainees and verified by one consultant neurosurgeon. \red{Inter-rater reliability was calculated using Cohen's Kappa, resulting in 0.949 (\gls{CI} 0.983–0.853) for the 6 general surgical aspects and 0.945 (\gls{CI} 0.981–0.842) for the 4 \gls{eTSA} specific aspects, as defined in the first and second column respectively under \Q{\gls{mOSATS} Assessment} in \fig{fig4}.} \fig{fig3} displays the \gls{mOSATS} distribution.

\section{Methods}

\subsection{Instrument segmentation and tracking}
\subsubsection{PRINTNet} 
The simplified diagram of the created architecture is displayed in the dashed green box of \fig{fig4}. The encoder is ResNet50 \cite{He2015}, with no pre-training: a well understood; strong performing; and lightweight \gls{CNN} commonly used for medical imaging tasks \cite{Wang2022}, particularly for \gls{eTSA} recognition \cite{Das2022}\cite{Das2023A}. The decoder is DeepLabV3 \cite{Chen2017}, commonly used in \gls{eTSA} segmentation \cite{Das2023B}, which utilises Atrous (also called dilation) convolutions, as opposed to skip connections found in other decoders. These convolutions skip a certain number of pixels (the dilation rate), which increases the receptive field without sacrificing spatial resolution or increasing the number of weights (and so computationally efficient), allowing object features to be captured on multiple spatial scales \cite{Chen2017}. This is particularly important for instrument segmentation in \gls{eTSA}, given the frequency in which instruments are entering and exiting the endoscopic view, and so the same instrument will be found in a variety of sizes.

\gls{SORT} begins with object detection using a \gls{CNN} as a feature extractor, followed by object estimating via velocity predictions, and finally ensuring the new objects detected and predicted trajectories of the old objects match \cite{Bewley2016}. DeepSORT extends \gls{SORT} through the use of a feature bank (storing features from previous frames), and matching these with the previous predictions \cite{Wojke2017}. StrongSORT extends DeepSORT through the use of an improved:  feature extractor; feature bank (now updater); velocity prediction algorithm; and matching algorithm \cite{Du2023}. Moreover, StrongSORT compensates for camera motion by estimating global rotation and translation between frames \cite{Du2023}, which is of importance for instrument tracking in \gls{eTSA}.  \gls{PRINTNet} utilises StrongSORT, replacing the object detection model with DeepLabV3. 

\subsubsection{Real-time implementation}
The implementation is done via the NVIDIA Holoscan SDK\footnote{https://github.com/nvidia-holoscan/holoscan-sdk} and runs on a NVIDIA Clara AGX\footnote{https://www.nvidia.com/en-gb/clara/intelligent-medical-instruments} \cite{Sinha2024}. The Holoscan SDK builds a TensorRT\footnote{https://developer.nvidia.com/tensorrt} engine, which optimises models through reductions in floating point precision; smaller model size; and dynamic memory allocation \cite{Sinha2024}.

\subsubsection{Metrics} 
\Gls{mIoU} was the evaluation metric for segmentation models. \gls{MOTP} was the evaluation metric for tracking models, and \gls{MOTA} is given as a secondary metric. \gls{MOTA} is calculated on every frame, and for frames where the ground-truth classification is unknown, it is assumed the ground-truth classification is unchanged since last known. \gls{MOTP} is calculated only on frames where ground-truth segmentations, and hence bounding-boxes, are known. \red{For these segmentation and tracking metrics a 100\% score indicates perfect overlap between the predicted and ground-truth annotation, with 0\% indicating no overlap or a missclassification.}

\gls{FPS} was the metric used to compare the speeds of the models. \red{A 25-\gls{FPS} model would match the native video frame rate and allow for real-time tracking, whereas a lower frame rate model would mean some frames in the video will be skipped.}

\subsubsection{Dataset split} 
4-fold cross-validation was implemented, as 15-videos is not sufficiently large for a reliable training to testing split. The folds were chosen such that each fold contained approximately the same number of images of a given instrument, but images from one video were only present in one fold. Five instrument classes (Blakesly; Irrigation Syringe; Retractable Knife; Dual Scissors; Surgical Drill) were removed from the analysis as they appeared in less than 4-videos, and so could not be present in each fold. This left four instrument classes (Blunt Dissector; Cup Forceps; Kerrisons; Pituitary Ronguers) as displayed in \fig{fig1}. \fig{fig2}a displays the stark data imbalance between the instrument classes. To mitigate the effect of overtraining on dominant classes, images of the Blunt Dissector and Kerrisons  were downsampled by 600 and 1200 respectively, and images of the Cup Forceps and Pituitary Ronguers were upsampled by 400. This was done per fold, and sampled images were chosen at random. \fig{fig2}b displays the resampled dataset (per fold).

\begin{figure*}[t]
\centering
\includegraphics[width=0.99\textwidth]{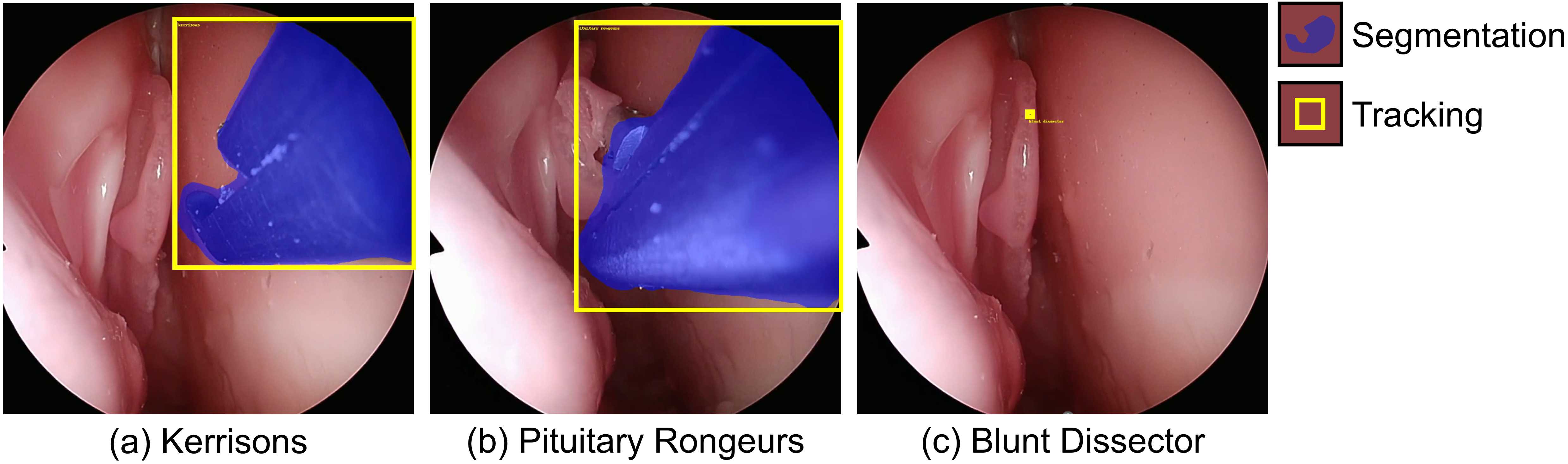}
\caption{Qualitative results of PRINTNet: (a) a strong example where the classification, segmentation, and tracking are accurate; (b) a common example where the classification and tracking are accurate, but the segmentation could be improved at the instrument tip; (c) an uncommon example where classification, segmentation, and tracking are all inaccurate. (See the Supplementary Material for the full video.)}
\label{fig5}
\end{figure*}

\begin{table*}[t]
\caption{Segmentation models' mIoU for each of the four instrument classes across the 4-folds. The highest mIoU for a given instrument is displayed in bold.}
\centering
\resizebox{0.99\textwidth}{!}{
\begin{tabular}{c|c|c|c|c|c|c}
Model & Blunt Dissector & Cup Forceps & Kerrisons & Pituitary Ronguers & All Instruments & No instrument \\ \hline
U-Net & 63.8±09.2 & 22.1±18.3 & 62.1±23.6 & 18.6±14.7 & 41.6±9.1 & 98.4±0.9 \\
SegFormer & 63.4±12.7 & \textbf{24.4±17.9} & 60.2±21.5 & 31.9±24.7 & 45.0±11.5 & 98.2±0.6 \\
\textbf{DeepLabV3} & \textbf{66.9±15.3} & 11.8±10.6 & \textbf{73.4±28.0} & \textbf{31.9±22.9} & \textbf{46.0±09.1} & \textbf{98.7±0.5} \\
\end{tabular}
}
\label{tab1}
\end{table*}

\subsubsection{Implementation details}
\red{To improve segmentation model training and generalisation, the following augmentation techniques were applied in sequence at random: horizontal flips; vertical flips; rotation; and colour jitters. As a compromise between having a sufficiently large batch size for finding optimal weights during gradient descent and a sufficiently high image resolution for meaningful feature extraction, models were training with a batch size of 16 with training images resized to $288\times512$ pixels\textsuperscript{2}, which was able to run on a single NVIDIA Tesla V100 Tensor Core 32-GB GPU.}   

\red{Cross-entropy was the loss function and Adam with learning rate 0.00006 was the optimiser, as these choices resulted in improved convergence over focal loss; and dice loss; and other optimiser variations. Each model was run for 50-epochs where the loss function was shown to be sufficiently small ($<0.04$) across all folds with minimal changes in subsequent epochs ($<0.005$ change after 100-epochs), and so restricting training to 50-epochs limits overfitting and reduces computational time. The model weights of the final (50\textsuperscript{th}) epoch was evaluated on the testing dataset with no early stopping procedure as to be a consistent choice which would not bias the model on any given fold.} 

The code is written in Python 3.8 using PyTorch 1.8.1 using CUDA 11.2, and is available at \red{https://github.com/dreets/printnet}. All videos and annotations are available at \red{https://doi.org/10.5522/04/26511049}.

\subsection{Surgical skill assessment}
In total, 34-metrics were extracted from the tracking data (see \fig{fig6}). In summary, it consisted of: time (e.g. instrument visible time); motion (e.g. acceleration); and usage metrics (e.g. number of instrument switches). 

For each metric, a \gls{PCC} was calculated against each \gls{mOSATS} aspect and summed \gls{mOSATS}. \red{A \gls{PCC} of 1.0 or -1.0 indicates direct positive or negative correlation respectively, with 0.0 indicating no correlation.}

Then, two classification tasks were then performed: multi-class \gls{mOSATS} (mean-averaged and rounded) and binary-class skill level (novice/expert). For each task, a Linear, \gls{SVM}; \gls{RF}; and \gls{MLP} model were trained, and boosted via \gls{ANOVA} feature selection. \red{A na\"{i}ve  classifier that only predicts the dominant class would achieve 33.3\% accuracy in multi-class by predicting \Q{3} and 66.7\% accuracy in binary-class by predicting \Q{novice}.}

\section{Results and Discussion}

\subsection{Instrument tracking and segmentation}
\subsubsection{Instrument segmentation} 
It is found that Blunt Dissector and Kerrisons are segmented well, with much worse performances for Cup Forceps and Pituitary Ronguers (see \tab{tab1}). This is due to the heavy data imbalance (\gls{mIoU}=0 for misclassifications), which is difficult to account for given the small number of images used for testing, even if balance sampling was implemented during training (see \fig{fig2}).

This difficulty in classification is likely because instrument handles are very similar, and take up a large portion of an image due to the image distortion, and so instruments must be distinguished by their relatively small tips. This can be more clearly seen in \fig{fig5}b where PRINTNet struggles in identifying the boundary of the Pituitary Rongeur, but is able to identify the boundary of Kerrisons (\fig{fig5}a), a dominant class. This again implies poor classification rather than poor segmentation, which is verified by ablation studies showing 82.2±0.2\% mIoU in binary segmentation.    

When compared to other segmentation models, DeepLabV3 has the highest overall \gls{mIoU}, although closely followed by SegFormer, which also has a significantly higher Cup Forceps \gls{mIoU}. Given more data, it is likely SegFormer will outperform DeepLabV3, as the transformer encoder performs better with larger datasets \cite{Sourget2023}, extracting both local and global spatial features \cite{Xie2021}. U-Net performs worse, as the skip connections between the \gls{CNN} encoder and upsampling decoder prevents derogation of local and not global spatial information \cite{Ronneberger2015}.

\begin{table}[t]
\begin{minipage}{0.99\columnwidth}
\caption{Tracking models' performance across the 4-folds. The highest value for a given evaluation metric is displayed in bold. Note detection frequency was set to 5.}
\centering
\resizebox{0.99\columnwidth}{!}{
\begin{tabular}{c|c|c|c}
Model & MOTP (\%) & MOTA (\%) & FPS (mean) \\ \hline
SORT & 59.1±03.1 & 77.9±07.1 & \textbf{24.7±00.8} \\
DeepSORT & 62.9±05.0 & 77.9±07.1 & 12.8±00.7 \\
\textbf{StrongSORT} & \textbf{71.9±05.5} & \textbf{77.9±07.1} & 10.6±02.9
\end{tabular}
}
\label{tab2}
\end{minipage}
\end{table}

\subsubsection{Instrument tracking} 
StrongSORT has the highest \gls{MOTP} as it accounts for camera motion, although at a lower \gls{FPS} when compared to SORT due to this extra computation (see \tab{tab2}). All models have an identical and high \gls{MOTA} as classification is determined by the same DeepLabV3 backbone.

Moreover, occasionally, \gls{PRINTNet} incorrectly predicts an instrument's classification; segmentation; and tracking; such as in \fig{fig5}c, caused by overpredicting the Blunt Dissector tracking paths from previous frames. These incorrect predictions increase the difficulty of surgical skill analysis as some metrics, such as time of instrument usage, may not be reliable.

\subsubsection{Real-time implementation}

\begin{figure*}[t]
\centering
\includegraphics[width=0.99\textwidth]{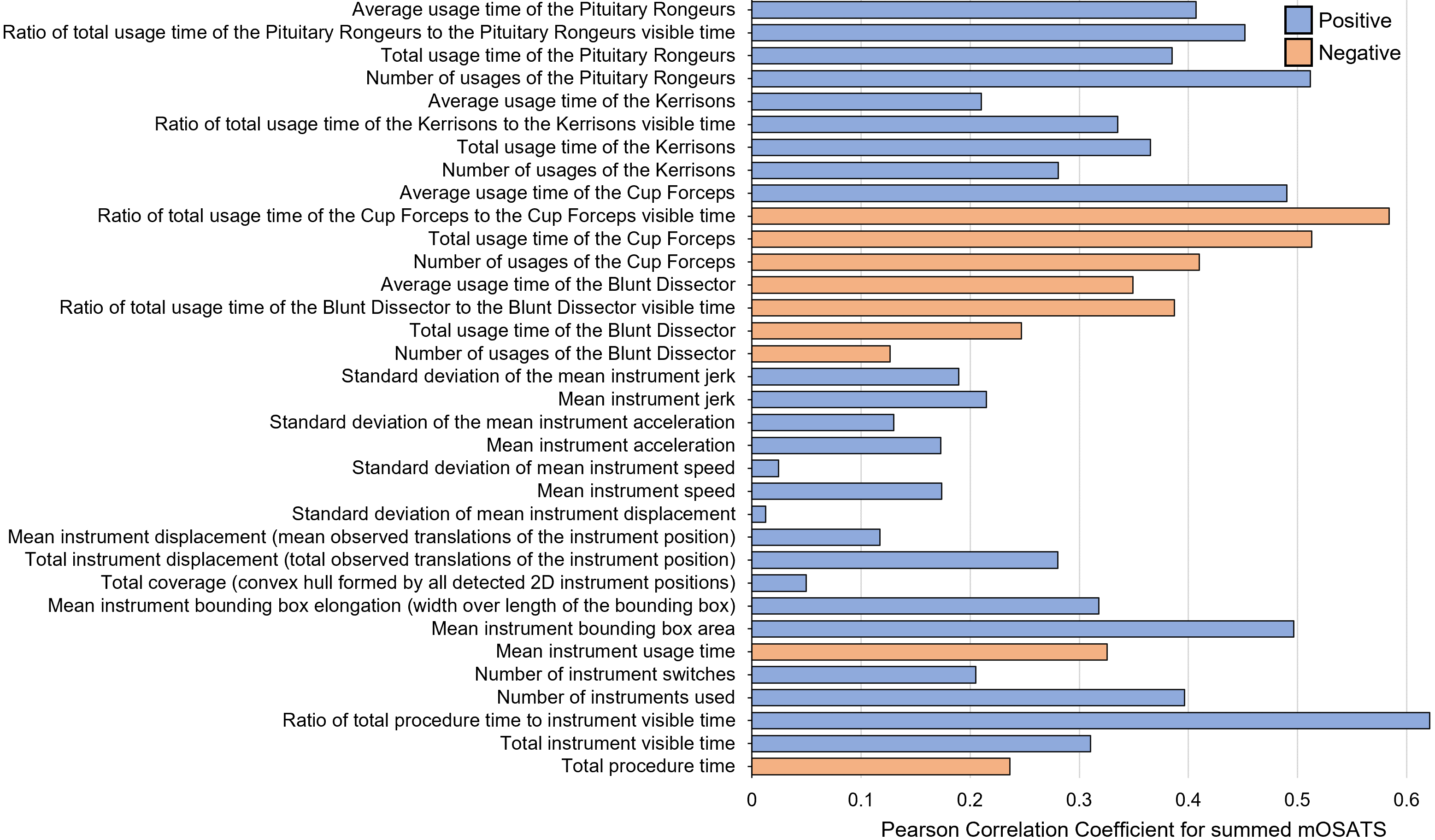}
\caption{Pearson Correlation Coefficient of the 34-metrics for summed mOSATS.}
\label{fig6}
\end{figure*}

\begin{table}[t]
\begin{minipage}{0.99\columnwidth}
\caption{Accuracy in surgical skill classification across the 4-folds. The highest value for a given metric is displayed in bold.}
\centering
\resizebox{0.99\columnwidth}{!}{
\begin{tabular}{c|c|c}
\multirow{2}{*}{Model} & Multi-class & Binary-class \\
 & mean mOSATS (\%) & skill level (\%) \\ \hline
Linear & \textbf{39.9±24.9} & 80.0±16.3 \\
Support Vector Machine & \textbf{46.7±26.7} & 80.0±26.7 \\
Random Forest & 40.0±38.9 & \textbf{73.3±24.9} \\
MultiLayer Perceptron & 26.7±24.9 & \textbf{86.7±16.3}
\end{tabular}
}
\label{tab3}
\end{minipage}
\end{table}

The accelerated \gls{PRINTNet} runs at 22-\gls{FPS} with a 100-millisecond delay at FP16 precision on the NVIDIA Clara AGX. This is sufficient for real-time use, so PRINTNet can be used during surgical training courses. (See Supplementary Material for a live demonstration of this setup.)

\subsection{Surgical skill assessment}
Distinguishing between expert and novice skill level achieved a high $87\%$ accuracy (see \tab{tab3}), in line with similar studies \cite{Lee2020}\cite{Fathollahi2022}\cite{Li2022}. However, there was poor accuracy in multi-class mean \gls{mOSATS} classification, although comparable to similar studies \cite{Li2022}. This highlights the complexity of the problem, with the implication that more data is required.  

Across the 10-aspects, time based metrics were stronger predictors than motion based metrics. This is seen in \fig{fig6} where \gls{PCC} for summed \gls{mOSATS} is shown. Specifically, \Q{ratio of total procedure time to instrument visible time} is found to be positively correlated with mOSATS, indicating instrument efficiency (i.e. a reduced idle time) is correlated with higher surgical skill. Interestingly, it is found the use of a Blunt Dissector or Cup Forceps is negatively correlated with mOSATS whereas Kerrisons and Pituitary Rongeurs are positively correlated.

The limited correlation between motion based metrics and mOSATS is an opposing result to that found in robotic thyroid surgery, where instrument motion in the absence of camera motion was a strong predictor \cite{Lee2020}. Removing this camera motion is tricky, as large endoscope movements are required to navigate through the nasal phase of \gls{eTSA} in order to get through the nostril (for both novice and expert surgeons), which outweighs the more subtle movements of the instruments. Although StrongSORT does compensate for this motion, more sophisticated models are needed.

\section{Conclusion}
Rating surgical skill via instrument tracking during minimally invasive surgery in an objective and reproducible manor remains a difficult task. Existing models have focused on real and robotic laparoscopic surgery, and these models have now been extended to simulated endoscopic surgery. 15-videos of the nasal phase of \gls{eTSA} were performed on a high-fidelity bench-top phantom during a training course and were recorded. They were later assessed for surgical skill by expert surgeons, and instruments were manually segmented. The created model, \gls{PRINTNet}, designed to classify; segment; and track the instruments during the nasal phase of \gls{eTSA} achieved 67\% and 73\% \gls{mIoU} for the dominant Blunt Dissector and Kerrisons classes, with 72\% \gls{MOTP}. 87\% accuracy was achieved with a \gls{MLP} when using the \gls{PRINTNet} tracking output to predict whether a surgeon was a novice or expert. Moreover, real-time speeds were achieved when run on a NVIDIA Clara AGX, allowing for real-time feedback for surgeons during training courses. This continuous monitoring of surgical skill allows novice surgeons to consistently improve their skill on simulated surgery before they are sufficiently skilled to perform real surgery. Future work will involve: modifying the model, such as with the use of temporal \cite{Fathollahi2022} or anchor free methods \cite{Ding2023}; collecting a larger dataset; and extending this work to real \gls{eTSA} - linking instrument tracking to both surgical skill and real patient outcomes. For now, this paper provides a new and unique publicly available dataset and baseline network, which can be improved on by the community.

\ack{
\red{
With thanks to Digital Surgery Ltd, a Medtronic company, for access to Touch Surgery Ecosystem for video recording, annotation and storage.
}
}

\fundingandinterests{\red{
This work was supported in whole, or in part, by the \gls{WEISS} [203145/Z/16/Z], the \gls{EPSRC} [EP/W00805X/1, EP/Y01958X/1, EP/P012841/1], the Horizon 2020 FET [GA863146], the Department of Science, Innovation and Technology (DSIT) and the Royal Academy of Engineering under the Chair in Emerging Technologies programme. Adrito Das is supported by the \gls{EPSRC} [EP/S021612/1]. Danyal Z. Khan is supported by a \gls{NIHR} Academic Clinical Fellowship and the \gls{CRUK} Pre-doctoral Fellowship. Hani J. Marcus is supported by \gls{WEISS} [NS/A000050/1] and by the \gls{NIHR} Biomedical Research Centre at \gls{UCL}. 
}
}

\bibliography{refs}
\end{document}